\documentclass[prc,superscriptaddress,unsortedaddress,twocolumn,showpacs,preprintnumbers,amsmath,amssymb,floatfix]{revtex4}
\usepackage[dvips]{graphicx}
\usepackage{amsmath}
\usepackage{amssymb}
\usepackage{times}
\usepackage{mathrsfs}
\usepackage{multirow}
\usepackage{bm}
\usepackage{wrapft}
\usepackage{color}

\def\ga{\mathrel{\mathpalette\fun >}}
\def\fun#1#2{\lower3.6pt\vbox{\baselineskip0pt\lineskip.9pt
\ialign{$\mathsurround=0pt#1\hfil##\hfil$\crcr#2\crcr\sim\crcr}}}

\def\ve{\varepsilon}

\def\non{\nonumber}

\newcommand{\beq}{\begin{equation}}
\newcommand{\eeq}{\end{equation}}
\newcommand{\bea}{\begin{eqnarray}}
\newcommand{\eea}{\end{eqnarray}}

\newcommand{\bfi}[1]{\mbox{\boldmath $#1$}}

\newcommand{\vk}{{\bfi k}}
\newcommand{\vp}{{\bfi p}}

\newcommand{\vrr}{{\bfi r}}

\newcommand{\vy}{{\bfi y}}

\begin{document}


\title{
New description of four-body breakup reaction}

\author{Takuma Matsumoto}
\email[]{tmatsumoto@nucl.sci.hokudai.ac.jp}
\affiliation{Meme Media Laboratory, Hokkaido University, Sapporo
060-8628, Japan} 

\author{Kiyoshi Kat\=o }
\email[]{kato@nucl.sci.hokudai.ac.jp}
\affiliation{Division of Physics, Graduate School of Science, Hokkaido
University, Sapporo 060-0810, Japan} 

\author{Masanobu Yahiro}
\email[]{yahiro@phys.kyushu-u.ac.jp}
\affiliation{Department of Physics, Kyushu University, Fukuoka 812-8581, Japan}

\date{\today}

\begin{abstract}
We present a novel method of smoothing discrete 
breakup cross sections calculated by 
the method of continuum-discretized coupled channels.  
The smoothing method based on the complex scaling method 
is tested with success for a
$^{58}$Ni($d$, $pn$) reaction at $80$~MeV as an example of 
three-body breakup reactions
and applied to a $^{12}$C($^6$He, $nn^4$He) reaction at 
$229.8$~MeV 
as an example of four-body breakup reactions.  
Fast convergence of the breakup cross section 
with respect to extending the model space is 
confirmed. The method is also applied to  
$^{12}$C($^6$He, $nn^4$He) and $^{208}$Pb($^6$He, $nn^4$He) reactions at 
$240$~MeV/$A$ and compared with the experimental data. 
\end{abstract}


\maketitle
Exploring unstable nuclei far from the stable line is 
one of the most important subjects in 
nuclear physics. The unstable nuclei have exotic properties such as 
the halo structure~\cite{Tanihata1,Tanihata2,Hansen} and the island of 
inversion~\cite{Warburton}. 
As a feature of reactions induced by unstable nuclei, 
the projectile easily breaks up into its constituents. 
One of the most reliable methods for treating the projectile 
breakup processes over a wide range of incident energies 
is the continuum-discretized coupled
channels (CDCC) method~\cite{CDCC-review1,CDCC-review2}. In CDCC, the 
scattering wave function of the total system is expanded with 
a finite number of bound and discretized continuum states 
of the projectile. 
The space spanned by these states is called the model space. 
The $S$-matrix elements calculated with CDCC converge as the
model space is extended~\cite{CDCC-convergence-1,CDCC-convergence-2}. 
The converged CDCC solution is the unperturbed solution of
the distorted Faddeev equations, and corrections to the solution are
negligible within the spatial region in which the breakup processes take 
place~\cite{CDCC-foundation1,CDCC-foundation2}. 

For scattering of a two-body projectile, the continuum states are
classified by linear and angular momenta, $k$ and $l$, respectively,
between the two constituents. In CDCC, these momenta are
taken up to upper limits, 
the $k$ continuum is divided into small bins and the
continuum states in each bin are averaged into a single state. This
discretization procedure is called the average (Av) method.
The Av method has been widely 
used, but its application has been limited to
three-body breakup reactions as we will show.
An alternative to the Av method is 
the pseudostate (PS) method~\cite{Matsumoto, Egami, Matsumoto3,
Matsumoto4, Moro1, THO-CDCC}, in which  
the continuum states $\{\psi^{(-)}(\bm{k})\}$ 
are replaced by pseudostates $\{\Phi_{n}\}$ 
obtained by diagonalizing the internal Hamiltonian of the projectile in
a space spanned by $L^2$-type basis functions. 
One can adopt the transformed harmonic oscillator
(THO)~\cite{Moro1} or the
Gaussian~\cite{Matsumoto,Egami} as the $L^2$-type basis functions.
The validity of the PS method was confirmed for 
scattering of two-body projectiles by the agreement between CDCC solutions 
calculated with the two discretization 
methods~\cite{Matsumoto, Egami, Matsumoto3, Matsumoto4}.

For scattering of $^6$He as a typical example of four-body breakup
reactions, CDCC with the PS method based on
Gaussian~\cite{Matsumoto3,Matsumoto4} or 
THO~\cite{THO-CDCC} basis functions was successful in describing
the elastic scattering at not only high 
energies but also low energies near the Coulomb barrier. 
Thus, the back-coupling effect of four-body breakup processes 
on the elastic scattering is well described by the PS method. 

For the three-body projectile, 
continuum wave functions 
$\psi^{(-)}(\bm{p},\bm{k})$ are classified  
by momenta ($\vp,\vk$) conjugate to two internal 
coordinates ($\vp$,$\vrr$) of the three-body system.
The breakup $S$-matrix elements calculated with CDCC, $S_{n}$, 
are discrete in $\vp$ and $\vk$, 
although the exact ones $S(\bm{p},\bm{k})$ are continuous. 
Thus, one needs a way of smoothing $S_{n}$. 
In principle this is possible by
calculating the smoothing factor 
$\langle\psi^{(-)}(\bm{p},\bm{k})|\Phi_{n}\rangle$~\cite{Matsumoto}, 
but in practice it is not easy, because 
evaluating $\psi^{(-)}(\bm{p},\bm{k})$ 
for many combinations of 
$\vp$, $\vk$ and $n$
is quite time consuming. 
Recently, we proposed two methods of obtaining the smoothing factor; 
one with a direct numerical
integration~\cite{LSCSM1} and another~\cite{LSCSM2} with the complex
scaling method (CSM)~\cite{ABC}. However, these require complicated
numerical calculations, so that  
the convergence of the smoothed breakup cross section with respect to 
increasing the model space is not sufficient particularly for 
the differential breakup cross section as a function of the excitation
energy $\ve$ of projectile, $d\sigma/d\ve$. 
If the Av method is applied with small momentum 
or energy bins, one can obtain 
$d\sigma/d\ve$ without evaluating the smoothing factor. 
This was done with the hyperradial  continuum  wave function~\cite{HRWF}, 
but the convergence of the CDCC solutions has not been
obtained yet for $d\sigma/d\ve$~\cite{{4body-CDCC-bin}}. 
A way of circumventing these difficulties is to construct 
a method of obtaining the smoothing factor without evaluating 
$\psi^{(-)}(\bm{p},\bm{k})$.

In this Rapid Communication, using CSM and CDCC, we propose a practical method 
of obtaining $d\sigma/d\ve$ as a continuous function of $\ve$ 
without evaluating 
$\psi^{(-)}(\bm{p},\bm{k})$. 
CSM is a powerful tool for obtaining 
many-body resonance and weakly bound states~\cite{CSM1}. 
Recently, it was applied to 
the electromagnetic transition of the core + nucleon + nucleon system, such as 
$^{6}$He and $^{11}$Li from the ground state to the continuum state with
$\ve$~\cite{CSM2,CSM3}. CSM is applicable not only for resonances 
but also for continuum states, so that 
the transition strength is obtained as a continuous function of $\ve$. 
The smoothing method proposed here is an alternative to 
the direct calculation of the smoothing factor 
$\langle\psi^{(-)}(\bm{p},\bm{k})|\Phi_{n}\rangle$ 
with the hyperradial continuum  wave function~\cite{Danilin1,Danilin2}.
The former is considered to be more practical than the latter, because 
the former does not require one to evaluate the continuum states 
$\psi^{(-)}(\bm{p},\bm{k})$ for many combinations of 
$\bm{p}$ and $\bm{k}$. 
The validity of the new method is
tested for a three-body breakup reaction, $^{58}$Ni($d$, $pn$) at 80~MeV,
in which the {\lq\lq}exact'' breakup cross section is obtainable 
by calculating the smoothing factor 
with the direct numerical integration. 
The new method is applied to the $^{12}$C($^6$He, $nn^4$He) reaction
at $229.8$~MeV. 
A merit of the present smoothing method is that one can see fast convergence 
of the calculated breakup cross section 
with respect to extending the model space. 
The method is also applied to $^{12}$C($^6$He, $nn^4$He) and 
$^{208}$Pb($^6$He, $nn^4$He) reactions at $240$~MeV/$A$ and compared with 
the experimental data. 
In principle, 
this method is applicable not only for four-body breakup reactions but also 
for many-body breakup reactions. 

We consider scattering of a projectile $B$ from a target $A$, in which 
$B$ is composed of three constituents ($x=b$, $c$ and $d$).  
The scattering is described by 
the four-body Schr\"odinger equation 
\begin{align}
(H-E_{\rm tot})|\Psi^{(+)}\rangle =0 
\label{eq:4b-Schr-ex}
\end{align}
with the outgoing boundary condition,
where the total energy $E_{\rm tot}$ satisfies 
$E_{\rm tot}=E_{\rm in}^{\rm CM}+\ve_0$  
for the corresponding incident energy $E_{\rm in}^{\rm CM}$ 
in the center of mass of the $B+A$ system
and the ground-state energy $\ve_0$ of $B$.
The total Hamiltonian $H$ of this system is defined by  
\begin{align}
 H &=K_R+U+H_{B}
\end{align}
with 
\begin{align}
 U &=U_{b}+U_{c}+U_{d}+
 V_{b}^{\rm Coul}+V_{c}^{\rm Coul}+V_{d}^{\rm Coul},\\ 
 H_{B}&=K_{y}+K_{r}+V_{bc}+V_{cd}+V_{db},
\end{align}
where $H_{B}$ is an internal Hamiltonian of $B$. 
The relative coordinate between $B$ and $A$ is denoted by 
$\bm{R}$, and  the internal coordinates of $B$ are denoted by 
a set of Jacobi coordinates $\bm{\xi}=(\bm{y},\bm{r})$. 
Momenta conjugate to coordinates $\bm{R}$ and $(\bm{y},\bm{r})$ are
represented by $\bm{P}$ and  
$(\bm{p},\bm{k})$, respectively. The kinetic energy
operator associated with $\bm{R}$ ($\bm{\xi}$) is represented
by $K_{R}$ ($K_\xi$), $V_{xx'}$
is a nuclear plus Coulomb interaction between $x$ and $x'$, 
and $U_{x}$ and 
$V_{x}^{\rm Coul}$ are nuclear and Coulomb potentials
between $x$ and $A$, respectively.  

In CDCC with the pseudostate discretization method, 
the scattering is assumed to take place 
in a model space~\cite{Matsumoto, Egami,
Matsumoto3, Matsumoto4}: 
\begin{eqnarray}
 {\cal P}=\sum_{n}|\Phi_{n}\rangle \langle \Phi_{n}|, 
 \label{eq:com-set} 
\end{eqnarray}
where ${\Phi}_{n}$ is an $n$th eigenstate obtained by 
diagonalizing $H_{B}$ with $L^2$-type basis functions. 
For simplicity, $B$ is assumed to have 
only one bound state ${\Phi}_{0}$. 
The four-body Schr\"odinger equation is then solved 
in the model space: 
\begin{align}
{\cal P}(H-E_{\rm tot}){\cal P}|\Psi^{(+)}_{\rm CDCC} \rangle =0.
\label{eq:4b-Schr}
\end{align}
The model space assumption has already been justified by the fact that 
calculated elastic and breakup cross sections 
converge with respect to extending the model space 
\cite{Matsumoto, Egami, Matsumoto3, Matsumoto4}.

The exact $T$-matrix element to a breakup state with $(\bm{p},\bm{k})$ 
can be described by 
\begin{eqnarray}
T_{\ve}(\bm{p},\bm{k},{\bm{P}})=
 \langle \psi^{(-)}_\ve(\bm{p},\bm{k}) \chi^{(-)}_\ve(\bm{P})|
U-V^{\rm Coul}_{B}
|\Psi^{(+)}\rangle
, \label{Tmat0}
 \label{exact-T}
\end{eqnarray}
where 
$V_{B}^{\rm Coul}$ is a sum of Coulomb interactions between $B$ 
and $A$ but the arguments are replaced by $R$: 
\begin{eqnarray}
 V_{B}^{\rm Coul}(R)=
V_{b}^{\rm Coul}(R)+V_{c}^{\rm Coul}(R)+V_{d}^{\rm Coul}(R). 
\end{eqnarray}
The final-state wave functions, $|\psi_\ve^{(-)}(\bm{p},\bm{k})\rangle$ and 
$|\chi_\ve^{(-)}(\bm{P})\rangle$, with the incoming boundary condition
are defined by  
\begin{eqnarray}
 \left[T_R+V^{\rm Coul}_B(R)-(E_{\rm tot}-\ve)\right]
  |\chi^{(-)}_\ve(\bm{P})\rangle &=&0,\\
 (H_{B}-\ve)|\psi_\ve^{(-)}(\bm{p},\bm{k}) \rangle&=&0, 
  \label{eq:Sch-B}
\end{eqnarray}
where 
$E_{\rm tot}-\ve=(\hbar P)^2/(2\mu_R)$
and
$\ve=(\hbar p)^2/(2\mu_{y})+(\hbar k)^2/(2\mu_{r})$ 
for reduced masses $\mu_{R}$ and $\mu_{\xi}$ of coordinates $\bm{R}$ and
$\bm{\xi}$, respectively. 
Inserting the approximately complete set Eq.
\eqref{eq:com-set} into Eq. \eqref{exact-T}, we can 
find \cite{Matsumoto, Egami, 
Matsumoto3, Matsumoto4} 
that the $T$-matrix element is well approximated by 
\begin{eqnarray}
T_{\ve}(\bm{p},\bm{k},{\bm{P}}) 
\approx \sum_{n \ne 0}
\langle \psi_\ve^{(-)}(\bm{p},\bm{k}) |\Phi_{n}\rangle
T_{n}
\label{approx-T}
\end{eqnarray}
with the CDCC $T$-matrix element
\begin{eqnarray}
T_{n}=\langle \Phi_{n}\chi^{(-)}_{\ve_n}(\bm{P}_n)|
 U-V^{\rm Coul}_{B}
 |\Psi^{(+)}_{\rm CDCC}\rangle  
\end{eqnarray}
to an $n$th discrete breakup state $\Phi_{n}$ with an eigenenergy
$\ve_n$.
Here Eq.~\eqref{approx-T} is derived by replacing $\bm{P}$ by $\bm{P}_n$
in $\chi^{(-)}_\ve(\bm{P})$.
$T_{n}$ is obtainable by CDCC, but 
it is quite hard to calculate 
the smoothing factor 
$\langle\psi_\ve^{(-)}(\bm{p},\bm{k})|\Phi_{n}\rangle $ directly with 
either numerical integration~\cite{LSCSM1} or CSM~\cite{LSCSM2}. 
Hence, we propose a new way of obtaining 
the differential cross section with respect to $\ve$ without 
calculating the smoothing factor. 

Using Eq. \eqref{approx-T}, one can rewrite 
the differential cross section as
\begin{eqnarray}
\frac{d^2\sigma}{d\ve d\Omega_{\bm{P}}} &=& 
\int d \vp' d \vk' 
\delta(\ve-\ve')
|T_{\ve'}(\vp',\vk',{\bm{P}'})|^2\non\\
&\approx& \frac{1}{\pi}{\cal R}(\ve,\Omega_{\bm{P}})
\label{xsec-1}
\end{eqnarray}
with the generalized response function 
\begin{eqnarray}
 {\cal R}(\ve,\Omega_{\bm{P}})
={\rm Im} \left(
\sum_{n,n'\ne 0} T_{n}^{*} 
\langle \Phi_{n}|G^{(-)}|\Phi_{n'}\rangle
T_{n'} 
\right),\label{response-fun}
\end{eqnarray}
where 
 $\displaystyle G^{(-)}=\lim_{\eta \to +0} (\ve-H_{B} - i\eta)^{-1}$.
In Eq. \eqref{response-fun}, there is no smoothing factor, as 
expected. Furthermore, the propagator $G^{(-)}$ operates only 
on spatially damping functions $\Phi_{n}$, so that the 
calculation of $\langle \Phi_{n}|G^{(-)}|\Phi_{n'}\rangle$ 
becomes feasible, as we will show.

CSM is now applied to evaluating 
$\langle \Phi_{n}|G^{(-)}|\Phi_{n'}\rangle$. 
The scaling transformation operator $C(\theta)$ and  
its inverse are defined by 
\begin{eqnarray}
\langle \vy,\vrr | C(\theta)|f \rangle 
&=& e^{3i\theta}f(\vy e^{i\theta},\vrr e^{i\theta}), \\
\langle f | C^{-1}(\theta)| \vy,\vrr \rangle 
&=& \left[ 
     e^{-3i\theta}f(\vy e^{-i\theta},\vrr e^{-i\theta})
     \right]^{*}.
\end{eqnarray}
Using the operators, one can get 
\begin{eqnarray} 
\langle \Phi_{n}|G^{(-)}|\Phi_{n'}\rangle
=\langle \Phi_{n}|C^{-1}(\theta) G_{\theta}^{(-)}C(\theta)|\Phi_{n'}\rangle ,
\label{C-G-1}
\end{eqnarray}
where 
\begin{eqnarray}
G_{\theta}^{(-)}=\lim_{\eta \to +0} \frac{1}{\ve-H^{\theta}_{B} - i\eta} . 
\end{eqnarray}
with 
$H_{B}^\theta=C(\theta)H_{B}C^{-1}(\theta)$.
When $-\pi<\theta<0$,
the scaled propagator $\langle\bm{\xi}|G^{(-)}_\theta|\bm{\xi}'\rangle$ 
is a damping function of $\bm{\xi}$ and $\bm{\xi}'$. 
It should be noted that although the scaling angle in
general calculations with CSM has been taken as positive, the angle in
the present situation becomes negative since 
$G^{(-)}$ has the incoming boundary condition. 
Hence, it can be expanded with $L^2$-type basis functions with high accuracy:
\begin{eqnarray}
G^{(-)}_\theta &\approx& \sum_i
  \frac{|\phi^\theta_i\rangle\langle\tilde{\phi}_i^\theta|}
  {\ve-\ve_i^\theta},
  \label{Gtheta2}
\end{eqnarray}
where $\phi_i^\theta$ is 
an $i$th eigenstate obtained by 
diagonalizing $H_{B}^\theta$
in a model space spanned by $L^2$-type basis functions: 
\begin{eqnarray}
\langle\tilde{\phi}_i^\theta|H_{B}^\theta
|\phi_{i'}^\theta\rangle&=
\ve_i^\theta \delta_{ii'} . 
\end{eqnarray}
By virtue of CSM, thus, we do not need to evaluate the
exact three-body continuum state $\psi_{\ve}^{(-)}(\vk,\vp)$ 
to obtain $\langle \Phi_{n}|G^{(-)}|\Phi_{n'}\rangle$. 

Inserting Eq. \eqref{Gtheta2} into Eq. \eqref{response-fun} by means of
Eq. \eqref{C-G-1} 
leads to a useful equation:
\begin{eqnarray}
 \frac{d^2\sigma}{d\ve d\Omega_{\bm{P}}}&\approx&
  \frac{1}{\pi}{\rm Im}\sum_i\frac{T_{i}^\theta \tilde{T}_i^\theta}
  {\ve-\ve_i^\theta}
  \label{sigma-CSM}
\end{eqnarray}
with 
\begin{eqnarray}
 \tilde{T}_i^\theta &\equiv&
 \sum_{n'} \langle \tilde{\phi}_i^\theta|C(\theta)  |\Phi_{n'} \rangle 
 T_{n'}, 
 \label{R-7} \\
  T_i^\theta&\equiv&
  \sum_n
  T_n^*   \langle \Phi_n|C^{-1}(\theta)|\phi_i^\theta \rangle.
  \label{R-8}
\end{eqnarray}
The principal result of this Rapid Communication
is that $C(\theta)$ and $C^{-1}(\theta)$ operate only on the
spatially damping function $\Phi_{n}$. This makes the calculation of 
$\langle \tilde{\phi}_i^\theta|C(\theta)  |\Phi_{n'} \rangle$ 
and $\langle \Phi_n|C^{-1}(\theta)|\phi_i^\theta \rangle$ feasible 
and makes possible the convergence of $d^2\sigma/d\ve d\Omega_{\bm P}$ 
with respect to extending the model space, as we will show. 
In other words, $C(\theta)$ and $C^{-1}(\theta)$ are not allowed to 
act on a nondamping function such as the plane wave, since 
the scaled function diverges asymptotically in this case. 

\begin{figure}[htbp]
\begin{center}
 \includegraphics[width=0.4\textwidth,clip]{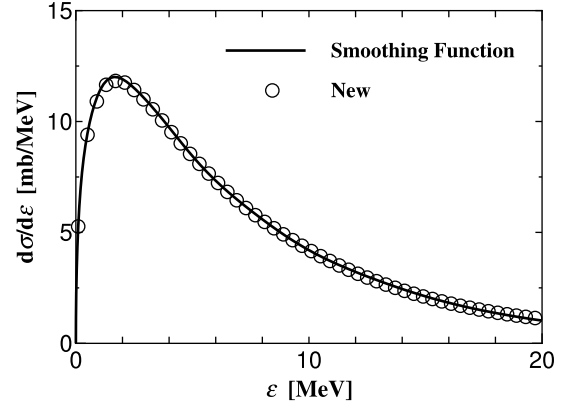}
 \caption{Calculated breakup cross section for $d$ + $^{58}$Ni scattering
 at 80 MeV. The solid line shows the result of the smoothing
 factor method; open circles show the result using the new method. } 
 \label{valid}
\end{center}
\end{figure}

We now test the validity of Eq. \eqref{sigma-CSM} for a three-body breakup 
reaction in which the ``exact'' 
breakup $T$-matrix element 
$T(\vk,\bm{P})=\sum_{n}\langle\psi^{(-)}(\bm{k})|{\Phi}_{n}\rangle{T_{n}}$
is obtainable 
by taking the overlap 
$\langle \psi^{(-)}(\bm{k})|{\Phi}_{n}\rangle$ directly with 
numerical integration. This approach is referred to here as 
the ``smoothing'' factor method~\cite{Matsumoto}.
As an example, we consider a $^{58}$Ni($d$, $pn$) reaction at 80 MeV; 
see Ref.~\cite{Matsumoto} for the details of the CDCC calculation. 
Figure~\ref{valid} shows the differential breakup cross
section $d\sigma/d\ve$ in which the double differential cross section of
Eq. \eqref{sigma-CSM} is integrated over the solid angle 
$\Omega_{\bm P}$ of  momentum ${\bm P}$. The new method (open circles)
yields the same result as the smoothing factor method (solid line). 
Thus, the new method is confirmed to be valid.

Next, the new method is applied to $^{12}$C$(^6$He, $nn^4$He$)$ 
scattering at 229.8 MeV as an example of a four-body breakup reaction; 
see Ref.~\cite{THO-CDCC} for optical potentials between $n$ and $^{12}$C 
and between $\alpha$ and $^{12}$C. 
As interactions $V_{nn}$ and $V_{n\alpha}$ in $H_{B}$, we take the
so-called GPT~\cite{GPT} and KKNN~\cite{KKNN} potentials, respectively. 
These potentials with a Gaussian form 
reproduce well data of low-energy nucleon-nucleon and 
nucleon-$^4$He scattering, respectively. 
The particle exchange between valence neutrons and neutrons 
in $^4$He is treated approximately with the orthogonality condition
model~\cite{OCM}.

For the diagonalization of $H_{B}$ and $H_{B}^\theta$, we adopt
the Gaussian expansion method (GEM)~\cite{H-Ka-Ki}. In GEM, 
the state of the $^4$He + $n$ + $n$ system is described by 
a superposition of three channels, each channel with 
a different set of Jacobi coordinates, $(\vy_c, \vrr_c)$.  
For each channel ($c$), 
the radial parts of the internal wave functions 
involving $\vy_c$ and $\vrr_c$ are 
expanded by a finite number of Gaussian basis functions 
\begin{eqnarray}
 \varphi_{j\lambda}(\bm{y}_c)&=&y_c^\lambda e^{-(y_c/\bar{y}_j)^2}
  Y_\lambda(\Omega_{y_c}), \non\\
 \varphi_{i\ell}(\bm{r}_c)&=&
  r_c^\ell e^{-(r_c/\bar{r}_i)^2}Y_\ell(\Omega_{r_c}),
\end{eqnarray}
respectively. Here $\lambda$($\ell$) 
is the angular momentum of $\bm{y}_c$($\bm{r}_c$), 
and the range parameters are taken to lie in geometric 
progression:
\begin{eqnarray}
 \bar{y}_j&=&(\bar{y}_{\rm max}/\bar{y}_{\rm 1})^
  {(j-1)/j_{\rm max}},\label{para1}\\
 \bar{r}_i&=&(\bar{r}_{\rm max}/\bar{r}_{\rm 1})^
  {(i-1)/i_{\rm max}}.\label{para2}
\end{eqnarray}
The parameters depend on $c$, but we omitted the dependence 
in Eqs. \eqref{para1} and \eqref{para2} for simplicity; 
see Ref.~\cite{Matsumoto3} for the details of 
the diagonalization and the definition of Jacobi coordinates.

\begin{table}[htbp]
 \caption{Gaussian range parameters.} 
 \begin{tabular}{c|ccccccc}
  \hline
  Set&$c$&$j_{\rm max}$&
  $\bar{y}_1$ (fm)&$\bar{y}_{\rm max}$ (fm)&$i_{\rm max}$
  &$\bar{r}_1$ (fm)&$\bar{r}_{\rm max}$ (fm)
  \\ \hline\hline
  \multirow{2}{*}{I} &3&10&   0.1&       10.0& 10&   0.5&  10.0   \\
  &1, 2&10&   0.5&       10.0&10&    0.5&  10.0 \\ \hline
 \multirow{2}{*}{II} &3 &15&   0.1&       20.0& 15&   0.5&  20.0   \\
  &1, 2&15 &  0.5&       20.0& 15&   0.5&  20.0 \\ \hline
  \multirow{2}{*}{III} &3 &20&  0.1&       50.0& 20&   0.5&  50.0   \\
  &1, 2&20&  0.5&       50.0&20&    0.5&  50.0 \\ \hline
 \end{tabular}
\label{basis-para}
\end{table}

In order to confirm the convergence of the breakup cross section 
with respect to extending the model space, we prepared three sets 
of basis functions shown in Table~\ref{basis-para}. 
For $0^+$ and ${1}^-$ states, 
maximum internal angular momenta are set to $\ell=\lambda=1$. 
For $2^+$ states, they are 
$\ell=\lambda=1$ for $c=1$ and 2, and $\ell=\lambda=2$ for $c=3$. 

\begin{figure}[htbp]
\begin{center}
 \includegraphics[width=0.4\textwidth,clip]{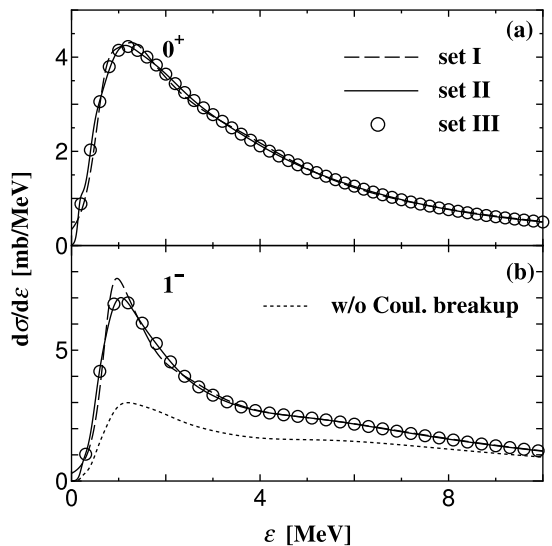}
 \caption{Convergence of the breakup cross sections to (a) $0^+$ continuum 
 and  (b) $1^-$ continuum. 
 Dashed lines, solid lines, and open circles correspond to
 results of sets I, II, and III, respectively. The dotted
 line in (b) shows the results when Coulomb breakup processes are
 switched off.}  
 \label{conv-Gtheta1}
\end{center}
\end{figure}

\begin{figure}[htbp]
\begin{center}
 \includegraphics[width=0.4\textwidth,clip]{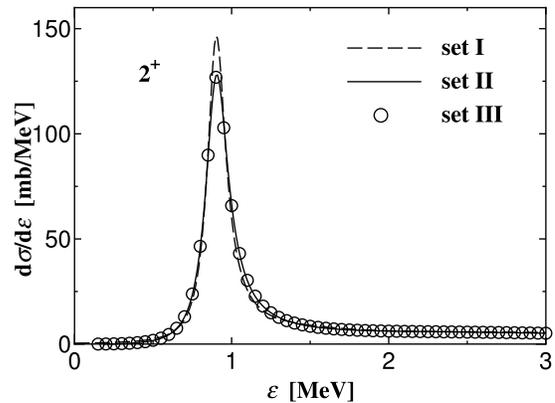}
 \caption{Convergence of the breakup cross sections to $2^+$ continuum. 
 Definition of lines is the same as in Fig.~\ref{conv-Gtheta1}. 
 }
 \label{conv-Gtheta2}
\end{center}
\end{figure}

\begin{figure}[htbp]
\begin{center}
 \includegraphics[width=0.4\textwidth,clip]{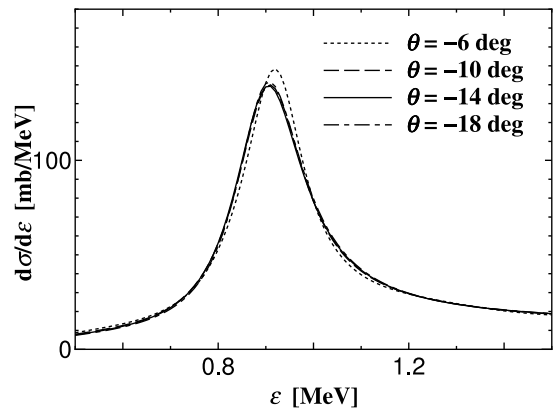}
 \caption{Dependence of the net breakup cross section 
 to $0^+$, $1^-$, and $2^+$ continua on $\theta$.}
 \label{conv-theta}
\end{center}
\end{figure}

Figure~\ref{conv-Gtheta1} shows the breakup cross sections $d \sigma/d\ve$ 
to $0^+$ and $1^-$ continua separately, and Fig.~\ref{conv-Gtheta2} 
shows the cross section to the $2^+$ continuum. 
For all the cross sections, sets II and III 
yield the same result, 
but the result of set I is somewhat different
from those of sets II and III. 
Thus, the convergence with respect to increasing the model space 
is obtained with set II. 
Figure~\ref{conv-theta} shows the $\theta$ dependence of the net 
breakup cross section to $0^+$, $1^-$, and $2^+$ continua
around the $2^+$ resonance peak. 
The net breakup cross section converges at $\theta=-14^\circ$
when $\theta$ decreases from $-6^\circ$ to $-18^\circ$. 

The present calculation includes Coulomb breakup processes that 
were neglected in the previous paper~\cite{Matsumoto4}. The effect 
is significant for the $1^-$ continuum, as shown by the dotted line in
Fig.~\ref{conv-Gtheta1}(b). It enhances the breakup cross section by a 
factor of 2 from the result without Coulomb breakup processes. 
The effect is negligible for $0^+$ and $2^+$ continua, 
although the latter is a dominant component of the breakup
cross section. Consequently, the Coulomb breakup effect in
the present reaction system is not significant for 
either the breakup reaction or elastic scattering.  
For heavy targets, Coulomb breakup processes dominate 
breakup reactions and Coulomb breakup reactions are a useful tool for
investigating properties of halo nuclei. The new method proposed here
can treat both nuclear and Coulomb breakup processes and then be used to
analyze their interference in the same framework. 

\begin{figure}[htbp]
\begin{center}
 \includegraphics[width=0.4\textwidth,clip]{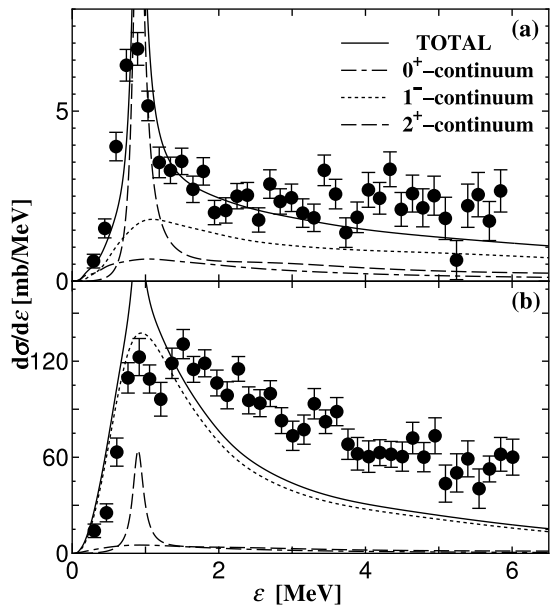}
 \caption{Comparison of the breakup cross section calculated by CDCC
 (solid line) with experimental data for (a) $^6$He + $^{12}$C 
 scattering 
 at 240 MeV/A and (b) $^6$He + $^{208}$Pb scattering at 240 MeV/$A$. 
 Dot-dashed, dotted, and dashed lines show contributions of $0^+$,
 $1^-$, and $2^+$ breakup, respectively. 
 The experimental data are taken
 from Ref.~\cite{Aumann}.} 
 \label{results}
\end{center}
\end{figure}

In Fig.~\ref{results}, the breakup cross section $d\sigma/d\ve$ calculated 
by the present method is compared with the experimental data for 
$^6$He + $^{12}$C and $^6$He + $^{208}$Pb reactions at 240
MeV/$A$~\cite{Aumann}.  
These data have already been analyzed by four-body distorted-wave Born
approximation (DWBA)~\cite{4BDW} 
and the eikonal approximation~\cite{4BEIK}. In 
the present calculation, we take the same potentials as in
Ref.~\cite{4BEIK} for $n$-$^{208}$Pb and $\alpha$-$^{208}$Pb systems. 
The optical potential for a $n$-$^{12}$C system is taken from the global
nucleon-nucleus potential~\cite{Koning}, while the optical potential for  
$\alpha$-$^{12}$C system is constructed from the $^{12}$C + $^{12}$C 
potential at 200 MeV/$A$~\cite{12C-12C} by changing the radius parameter 
from $^{12}$C to $\alpha$. 
Nuclear breakup is dominant for $^6$He + $^{12}$C 
scattering at 240 MeV/A, while Coulomb breakup 
to the $1^-$ continuum is dominant for $^6$He + $^{208}$Pb scattering. 
For a $^{12}$C target, the present theoretical result is consistent 
with the experimental data
except for the peak of the 2$^+$ resonance around 
$\ve=1$ MeV. 
This overestimation is also seen in the results of four-body DWBA, and 
the problem is partly solved by considering the experimental energy
resolution. 
For $^{208}$Pb target, 
the present method underestimates the experimental data 
at $\ve \ga 2$~MeV. A possible origin of this underestimation is that 
the inelastic breakup reactions are not included in the present calculation. 
{As mentioned in Ref.~\cite{4BDW}, 
the inelastic breakup
effect is not negligible, and the elastic breakup cross 
section calculated with four-body DWBA 
also underestimates the data.

In summary, we have proposed a practical method of calculating 
the differential breakup cross section as a continuous function of 
the excitation energy of a projectile, by combining CDCC
and CSM.
This method does not require one to {calculate the continuum
wave functions} of the projectile. 
All we have to do is just diagonalize 
the projectile Hamiltonian and the scaled Hamiltonian 
with $L^2$-type basis functions. 
In the present formalism, the scaling operator $C(\theta)$ operates only 
on spatially damping functions and hence 
the differential breakup cross section converges quickly as the model
space is extended.  
The method is successful in reproducing the data on 
$^6$He + $^{12}$C and $^6$He + $^{208}$Pb reactions at
240 MeV/$A$. 
In principle, the present
formalism is applicable for many-body breakup reaction, if the
diagonalization of the projectile Hamiltonian and the scaled Hamiltonian
is feasible.


The authors would like to thank Dr. K. Ogata for useful
and helpful discussions. MY is grateful to M.~Rodr\'{i}guez-Gallardo
and A.~M.~Moro for discussions and suggestions at DREB2009. The
numerical calculations of this work were performed on the computing
system at the Research Institute for Information Technology of Kyushu
University.


\end{document}